\documentclass[pra,
,tightenlines
,showkeys
,showpacs
,twocolumn
]{revtex4-2}

\usepackage{amsmath,amssymb}
\usepackage{bm}
\usepackage[dvipdfmx]{graphicx}
\usepackage{ascmac}
\usepackage{fancybox}
\usepackage{float}
\usepackage{enumerate}
\usepackage{color}
\usepackage{amsthm}
\usepackage{mathrsfs}
\usepackage{comment}
\allowdisplaybreaks[1]
\newtheorem*{th.}{Theorem}

\newcommand{\ket}[1]{|#1\rangle}

\usepackage{ulem}

\begin{document}

\title{Suppressing Detuning-Induced Bias in Ramsey Magnetometry with Composite Pulses}

\author{Shingo Kukita}
\email{kukita@nda.ac.jp}
\affiliation{Department of Computer Science, National Defense Academy of Japan, Yokosuka, Kanagawa, 239-0811, Japan}
\author{Kento Nawata}              
\affiliation{Department of Electrical, Electronic, and Communication Engineering, Faculty of Advanced Science and Engineering, Chuo University, Kasuga, Bunkyo-ku, Tokyo 112-8551, Japan}
\author{Yuichiro Matsuzaki}
\email{ymatsuzaki872@g.chuo-u.ac.jp}
\affiliation{Department of Electrical, Electronic, and Communication Engineering, Faculty of Advanced Science and Engineering, Chuo University, Kasuga, Bunkyo-ku, Tokyo 112-8551, Japan}

\begin{abstract}
Quantum sensing estimates a physical parameter encoded in the state of a probe; with
independent spin probes the precision follows the standard quantum limit. 
Studies of sensing precision often assume that the parameters entering the model, such as
the noise, are known. In practice these parameters are not always known, and a mismatch
between the assumed and actual values induces a systematic error.
Here we study single-qubit Ramsey magnetometry of a DC magnetic field under an unknown detuning between
the actual and nominal spin frequencies:
A first pulse puts the qubit into a superposition of its two states,
the field to be sensed then adds a relative phase during an exposure stage, and a second pulse enables
the readout.
In our setting, the field acts only during the exposure stage, whereas the detuning acts throughout the whole protocol.
We analyze how the detuning biases the estimate, 
preventing
the total estimation error from
following the standard quantum limit.
We then construct a composite-pulse preparation and readout that exploits the difference in 
the intervals over which the field and the detuning act to cancel the detuning to first order.
We evaluate the
performance of this composite-pulse protocol and show that it suppresses the
detuning-induced bias.
\end{abstract}
\maketitle

\section{Introduction}
\label{sec:intro}
Precise measurement underlies much of modern science and
technology~\cite{degen2017quantum}. Magnetic-field sensing in particular finds broad
application across physics, materials science, chemistry, and
biology~\cite{barry2020sensitivity,aslam2023quantum,montenegro2025quantum}. A typical
magnetometer encodes the field in the phase accumulated by spin probes: after preparing a
superposition, exposing the spins to the field for a fixed time, and reading out the
resulting phase, one infers the field
strength~\cite{maze2008nanoscale,taylor2008high,wolf2015subpicotesla}. With $M$
independent, identically prepared spins the standard deviation of the estimate scales as
$\mathcal{O}(M^{-1/2})$, the standard quantum limit
(SQL)~\cite{giovannetti2004quantum,huang2024entanglement}.

The performance of these protocols depends on how well the model of the experiment is known.
When the noise parameters are known, their effect can be corrected, and averaging over many
measurements reduces the statistical error toward the standard quantum limit. In practice
the model may contain parameters that are not always known~\cite{suzuki2020quantum}, such as an imperfectly
characterized noise rate or an imperfection in the coherent control like an over-rotation or
an unknown detuning~\cite{sugiyama2015precision,takeuchi2019quantum}. The correction is then inexact,
and the resulting systematic error keeps the total estimation error from following the standard quantum limit.

Several strategies have been proposed to reduce systematic errors in metrology. In weak
measurement under decoherence, postselection can lower the error of the
estimate~\cite{pang2016protecting}. Measurements in the Zeno regime have also been used to
suppress such errors~\cite{shimada2021quantum}. A complementary line of work applies quantum
error mitigation, which reduces bias by post-processing the measurement data, for example
through virtual
purification~\cite{PhysRevLett.129.250503,PhysRevA.109.022410,kwon2025virtual,lin2025error}.
These approaches act after the measurement rather than at the level of the control.

More recently, Ref.~\cite{vwpw-7rlb} analyzed the systematic error from detuning, a
deviation of the drive frequency from the spin resonance, for the GHZ-based sensing scheme
of Refs.~\cite{PhysRevLett.115.170801,jones2009magnetic}, and introduced a composite-pulse
sequence to compensate the detuning. Composite pulses were developed in nuclear magnetic
resonance~\cite{levitt1986composite,counsell1985analytical,PhysRevA.67.042308,Levitt2008}
and are widely used for robust quantum
control~\cite{PhysRevA.77.052334,Ball_2021,Zanon-Willette_2018}; a recent study also applies
them to sensing in the presence of pulse errors~\cite{lancaster2025quantum,aiello2013composite}. 


Here we propose a way to
suppress the detuning-induced bias in single-qubit Ramsey magnetometry
\cite{ramsey1950molecular} with composite pulses. In this protocol, a first pulse puts the
qubit into a superposition, the field to be sensed adds a phase during the exposure, and a second pulse
converts this phase into a population difference for readout.
In our setting, the field acts only during the exposure, whereas the detuning acts throughout the protocol. Focusing on the estimation of a static (DC) magnetic field, we analyze how the detuning biases the estimate,
preventing the total estimation error from following the standard quantum limit.
We then design a composite-pulse preparation and readout that exploits the difference in the intervals over
which the field and the detuning act to cancel the detuning to first order. We evaluate the
performance of the resulting composite-pulse protocol under depolarizing noise.

The paper is organized as follows. Section~\ref{sec:setup} sets up the single-qubit Ramsey
protocol and introduces the detuning error and depolarizing noise.
Section~\ref{sec:protocols} analyzes the conventional Ramsey protocol, which we call a single-pulse protocol, shows that the detuning
imposes an error floor, and constructs the composite-pulse preparation and readout that
removes the first-order detuning. Section~\ref{sec:performance} evaluates the performance of
both protocols. Section~\ref{sec:summary} summarizes the results and discusses extensions.

\section{Sensing model}
\label{sec:setup}
We consider magnetic-field sensing with a single qubit.
We first set up the ideal protocol and add depolarizing noise. 
We then study the systematic error caused by detuning, which introduces a systematic bias.

\subsection{Noisy sensing}
\label{sec:setup-protocol}
The sensor is a single qubit, whose state space is spanned by $\{\ket{0},\ket{1}\}$.
On this space the Pauli matrices $\sigma_{x},\sigma_{y},\sigma_{z}$ act in the standard way, with $\sigma_{0}$ denoting the identity.
We fix the convention $\sigma_{z}\ket{0}=\ket{0}$ and $\sigma_{z}\ket{1}=-\ket{1}$,
which together with the standard commutation relations determines the action of $\sigma_{x}$ and $\sigma_{y}$.

The protocol comprises three steps: preparation, exposure, and readout.
We initialize the qubit in $\ket{0}$ and apply a $\pi/2$ pulse, implemented by an AC driving
field, to prepare an equal superposition of $\ket{0}$ and $\ket{1}$. The drive is then turned off, and the qubit is exposed to the static (DC) magnetic field to be sensed. The field shifts the qubit's transition frequency by an amount $\Omega$ proportional to the field strength. This shift is described by the exposure Hamiltonian
\begin{equation}
\tilde{H}_{\mathrm{mag}}=\frac{\Omega}{2}\sigma_{z}.
\label{eq:Hmag}
\end{equation}
During an exposure time $t$, the state accumulates a relative phase $\Omega t$ between
$\ket{0}$ and $\ket{1}$.
A second $\pi/2$ pulse converts this phase into a population difference, which we read out by
measuring in the $\{\ket{0},\ket{1}\}$ basis.

In the ideal case---no detuning and no depolarizing noise---the probability of the
outcome $\ket{0}$ is
\begin{equation}
P_{0}=\frac{1}{2}+\frac{1}{2}\sin(\Omega t)\approx\frac{1}{2}+\frac{1}{2}\Omega t,
\qquad \Omega t\ll 1.
\label{eq:P0-ideal}
\end{equation}
The field is thus encoded in the outcome probability $P_{0}$. We restrict attention
to the regime $\Omega t\ll1$ of a small accumulated phase, in which $P_{0}$ is approximately
linear
in $\Omega t$.

To estimate $\Omega$, we repeat the protocol over $M$ independent trials.
In the $j$th trial we let $s_{j}$ be the indicator of the outcome $\ket{0}$ (i.e.\ $s_{j}=1$
if the outcome is $\ket{0}$, and $s_{j}=0$ otherwise);
each $s_{j}$ is then a Bernoulli variable with mean $P_{0}$.
We record the empirical frequency $S=\sum_{j=1}^{M}s_{j}/M$,
which converges to $P_{0}$ as $M\to\infty$, and invert Eq.~\eqref{eq:P0-ideal} to obtain the estimator
\begin{equation}
\tilde{\Omega}=\frac{\arcsin(2S-1)}{t}\approx\frac{2}{t}\left(S-\frac{1}{2}\right),
\label{eq:estimator-ideal}
\end{equation}
where the second expression is the linearization valid for $\Omega t\ll1$.
In this regime the estimator is unbiased, $\langle\tilde{\Omega}\rangle=\Omega$.
We quantify the performance by the mean-square error, obtained from the binomial fluctuations of
$S$,
\begin{equation}
E_{\mathrm{MS}}=\bigl\langle(\tilde{\Omega}-\Omega)^{2}\bigr\rangle
=\frac{4}{t^{2}}\frac{P_{0}(1-P_{0})}{M}\approx\frac{1}{M t^{2}},
\label{eq:mse-ideal}
\end{equation}
where $P_{0}(1-P_{0})=\tfrac14+\mathcal{O}((\Omega t)^{2})$.
Its square root scales as $\mathcal{O}(M^{-1/2})$, the SQL.

We now include depolarizing noise. At rate $\gamma$, over a time $\tau$ the depolarizing
channel acts as
\begin{equation}
\rho\;\longrightarrow\;e^{-\gamma \tau}\rho+\bigl(1-e^{-\gamma \tau}\bigr)\frac{\sigma_{0}}{2},
\label{eq:depol}
\end{equation}
driving $\rho$ toward the maximally mixed state $\sigma_{0}/2$. The qubit is subject to
this noise continuously, during both the pulses and the exposure.

With depolarizing noise the outcome probability becomes
\begin{equation}
P_{0}^{(\gamma)}=\frac{1}{2}+\frac{1}{2}e^{-\gamma\tau}\sin(\Omega t),
\label{eq:P0-gamma}
\end{equation}
where the $\sin(\Omega t)$ term now carries a factor $e^{-\gamma\tau}$.
Here $\tau$ is the total duration of one trial,
the sum of the exposure time $t$ and the durations of the two pulses. We take the depolarizing rate $\gamma$ to be known.
The damping factor $e^{-\gamma\tau}$ is then known, and we rescale the estimator of
Eq.~\eqref{eq:estimator-ideal} to undo it,
\begin{equation}
\tilde{\Omega}=\frac{1}{t}\arcsin\!\left(\frac{2S-1}{e^{-\gamma\tau}}\right)
\approx\frac{2}{t}\,\frac{S-\tfrac12}{e^{-\gamma\tau}},
\label{eq:estimator}
\end{equation}
where the second expression is the linearization for $\Omega t\ll1$.
The linear estimator remains unbiased, $\langle\tilde{\Omega}\rangle=\Omega$.

The rescaling in Eq.~\eqref{eq:estimator} carries a cost:
it multiplies the ideal mean-square error of Eq.~\eqref{eq:mse-ideal} by $e^{2\gamma\tau}$,
\begin{equation}
E_{\mathrm{MS}}=\bigl\langle(\tilde{\Omega}-\Omega)^{2}\bigr\rangle
\approx\frac{e^{2\gamma\tau}}{M t^{2}}.
\label{eq:mse-gamma}
\end{equation}
This prefactor is the cost of undoing the damping.
It raises the error but leaves the $\mathcal{O}(M^{-1/2})$ scaling unchanged.
This is unlike entanglement-enhanced metrology~\cite{huelga1997improvement},
where decoherence degrades the Heisenberg scaling toward the SQL.

\subsection{Detuning error}
\label{sec:setup-detuning}
We now specify the $\pi/2$ pulses introduced above in detail.
In the absence of control the qubit Hamiltonian is $H_{0}=(\omega_{0}/2)\sigma_{z}$,
where $\omega_{0}$ is the nominal transition frequency.
A drive adds $H_{1}(s)=(\Lambda/2)\cos(\omega s+\phi)\sigma_{x}$, with time $s$, strength $\Lambda$,
frequency $\omega$, and phase $\phi$.
We take the resonant case $\omega=\omega_{0}$ and the regime $\omega_{0}\gg\Lambda$.
We then move to the frame rotating at $\omega_{0}$ and apply the rotating-wave approximation (RWA).
This reduces the controlled evolution to a single-axis rotation,
\begin{align}
\tilde{U}(\theta,\phi)=&\cos\frac{\theta}{2}\,\sigma_{0}
-i\sin\frac{\theta}{2}\,\vec{n}_{\phi}\cdot\vec{\sigma},\nonumber\\
\vec{n}_{\phi}=&(\cos\phi,\sin\phi,0),
\label{eq:rotation}
\end{align}
where $\theta=\lambda s$ with $\lambda=\Lambda/2$ is the rotation angle and $\vec{\sigma}=(\sigma_{x},\sigma_{y},\sigma_{z})$.
The derivation is given in Appendix~\ref{app:rwa}.
Unless stated otherwise, we set $\lambda=1$ and measure all times in units of the pulse
duration, so the angle $\theta$ equals the duration $s$.

Note that the exposure is described in the same rotating frame.
In the laboratory frame the $\sigma_{z}$ coefficient during the exposure is $(\omega_{0}+\Omega)/2$, combining the nominal splitting and the target field.
As $\sigma_{z}$ commutes with the frame rotation, moving to the rotating frame removes the $\omega_{0}$ part and leaves the exposure Hamiltonian $\tilde{H}_{\mathrm{mag}}$ of Eq.~\eqref{eq:Hmag}, where the tilde denotes this frame.

The actual transition frequency departs from the nominal value $\omega_{0}$ by an
unknown but constant detuning $\delta$, becoming $\omega_{0}+\delta$, while the
control remains referenced to $\omega_{0}$.
Unlike the target field $\Omega$, which acts only during the exposure, the detuning is present throughout the protocol.
It shifts $\Omega\to\Omega+\delta$ in the exposure Hamiltonian of Eq.~\eqref{eq:Hmag},
and modifies each pulse as
\begin{align}
\tilde{U}_{\delta}(\theta,\phi)=&\cos\frac{R\theta}{2}\,\sigma_{0}
-i\sin\frac{R\theta}{2}\,\vec{n}_{\phi,\delta}\cdot\vec{\sigma},\nonumber\\
\vec{n}_{\phi,\delta}=&\frac{(\cos\phi,\sin\phi,\delta)}{R},
\end{align}
with $R=\sqrt{1+\delta^{2}}$.
Expanding to first order in $\delta$,
\begin{equation}
\tilde{U}_{\delta}(\theta,\phi)\approx
\tilde{U}(\theta,\phi)-i\,\delta\sin\frac{\theta}{2}\,\sigma_{z}
+\mathcal{O}(\delta^{2}).
\label{eq:Udelta}
\end{equation}

\section{Detuning bias and its suppression}
\label{sec:protocols}

The detuning of Sec.~\ref{sec:setup-detuning} biases the estimate, and with naive control
pulses this bias survives as a floor on the mean-square error. To overcome this floor, we
propose a composite-pulse protocol that cancels the detuning to first order in $\delta$. We
replace each single pulse of the preparation and readout by a tailored sequence of pulses,
whose individual detuning contributions are arranged to cancel, so the sequence reproduces
the intended rotation while removing the first-order detuning. We first analyze the
single-pulse protocol and derive the floor, and we then construct the proposed
composite-pulse protocol.

\begin{figure*}[t]
\centering
\includegraphics[width=\textwidth]{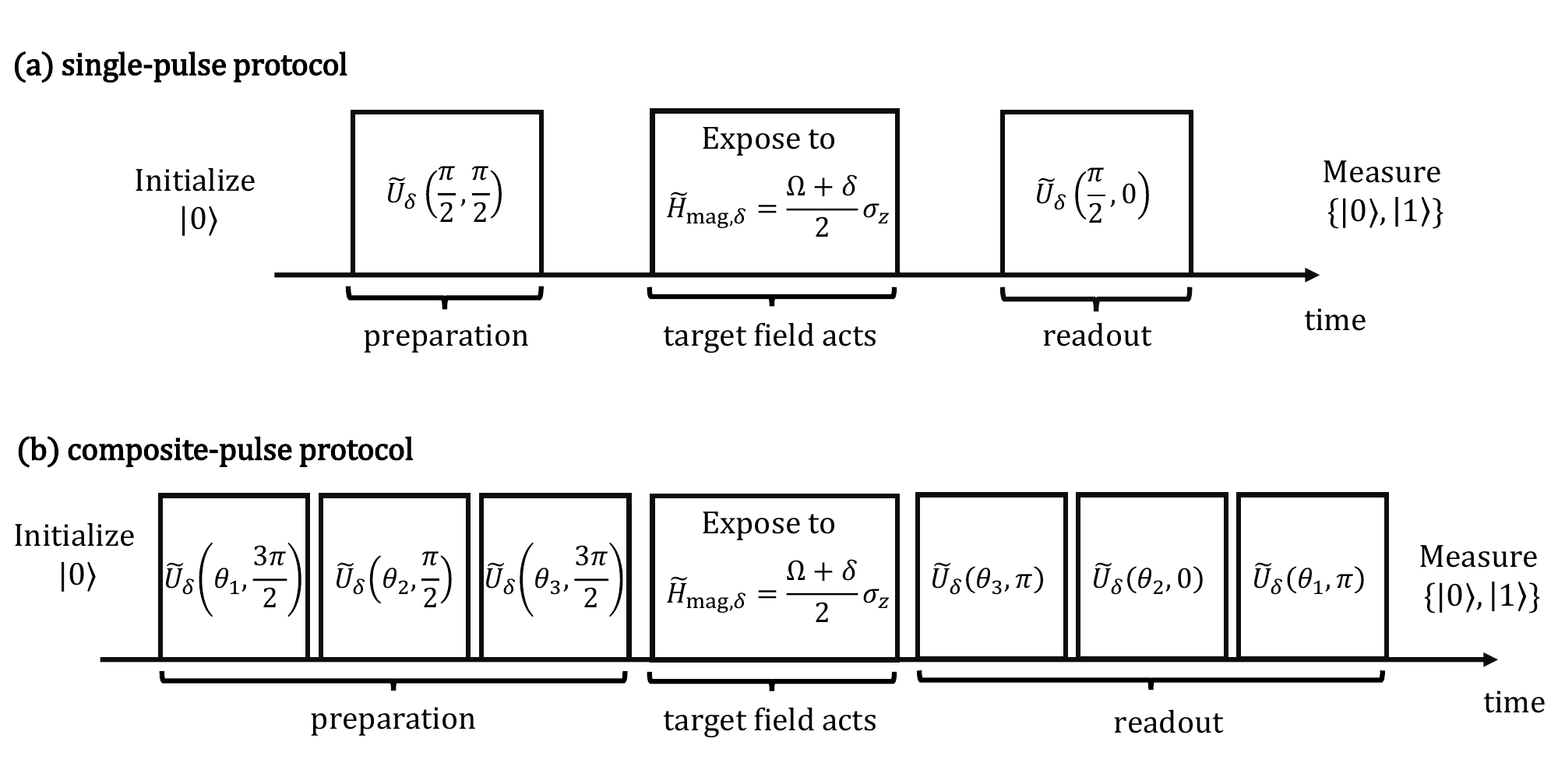} 
\caption{Schematic of the two protocols. (a) In the single-pulse protocol, a $\pi/2$ pulse
prepares an equal superposition, the qubit is exposed to the field for a time $t$, and a
second $\pi/2$ pulse maps the accumulated phase onto the populations, which are read out in
the $\{\ket{0},\ket{1}\}$ basis. (b) In the composite-pulse protocol, each $\pi/2$ pulse is
replaced by a three-pulse sequence with areas $\theta_{1},\theta_{2},\theta_{3}$ and phases
chosen to cancel the detuning to first order.}
\label{fig:example}
\end{figure*}

\subsection{Single-pulse protocol}
\label{sec:conventional}
We implement each preparation and readout of the protocol of Sec.~\ref{sec:setup-protocol} as a single pulse,
Eq.~\eqref{eq:rotation} with rotation angle $\pi/2$.
We call this the single-pulse protocol~[Fig.~\ref{fig:example}(a)].
As described in Sec.~\ref{sec:setup-detuning}, the detuning shifts $\Omega\to\Omega+\delta$ during the exposure and replaces each pulse by its detuned form $\tilde{U}_{\delta}$ of Eq.~\eqref{eq:Udelta}.
This degrades the sensing performance.
We quantify the degradation to first order in $\delta$.

We trace the state evolution through the protocol to first order in $\delta$ and
$\Omega$. The preparation pulse $\tilde{U}_{\delta}(\pi/2,\pi/2)$ rotates $\ket{0}$
about the $y$ axis, producing the superposition
\begin{equation}
\ket{0}\;\longrightarrow\;\frac{1}{\sqrt{2}}\bigl[(1-i\delta)\ket{0}+\ket{1}\bigr].
\end{equation}
The exposure $e^{-i\tilde{H}_{\mathrm{mag},\delta}t}$, with
$\tilde{H}_{\mathrm{mag},\delta}=(\Omega+\delta)\sigma_{z}/2$, imprints the relative
phase $(\Omega+\delta)t$, giving
\begin{equation}
\ket{\psi_{1}}=\frac{1}{\sqrt{2}}\Bigl[\bigl(1-i\delta-\tfrac{i}{2}(\Omega+\delta)t\bigr)\ket{0}
+\bigl(1+\tfrac{i}{2}(\Omega+\delta)t\bigr)\ket{1}\Bigr].
\end{equation}
The readout pulse $\tilde{U}_{\delta}(\pi/2,0)$ rotates about the $x$ axis, giving the
final state
\begin{equation}
\ket{\psi_{2}}=\frac{1}{2\sqrt{2}}\Bigl[\bigl(2+\Omega t+\delta t+2(1-i)\delta\bigr)\ket{0}
+\bigl(2-\Omega t-\delta t-2\delta\bigr)\ket{1}\Bigr],
\end{equation}
up to a global phase. The probability of the outcome $\ket{0}$ is then
\begin{equation}
P_{0,\delta}=\bigl|\langle 0|\psi_{2}\rangle\bigr|^{2}
=\frac{1}{2}+\frac{1}{2}\Omega t+\frac{1}{2}\delta t+\delta
+\mathcal{O}(\delta^{2},\Omega^{2},\Omega\delta).
\label{eq:P0-detuned}
\end{equation}
The term $\tfrac{1}{2}\delta t$ is the detuning accumulated during the exposure. The
constant term $\delta$ comes from the two pulses. As the depolarizing channel is
isotropic, it commutes with any unitary. The whole-trial noise then collects into a
single factor $e^{-\gamma\tau}$,
\begin{equation}
P_{0,\delta}^{(\gamma)}\simeq \frac{1}{2}
+\frac{1}{2}e^{-\gamma\tau}\bigl(\Omega t+\delta t+2\delta\bigr),
\label{eq:P0-detuned-gamma}
\end{equation}
where $\tau=t+\pi$ is the time of one trial: the exposure time $t$ plus the two
$\pi/2$ pulses, $\pi$ in total.

We here take the simple case in which $\delta$ is common to all $M$ trials.
A detuning that fluctuates from trial to trial modifies the analysis slightly, which we
consider later. As the detuning is unknown, we estimate as if $\delta=0$ and apply the
corrected estimator of Eq.~\eqref{eq:estimator}. Substituting
$\langle S\rangle=P_{0,\delta}^{(\gamma)}$ into its linearization gives the bias
\begin{equation}
b=\langle\tilde{\Omega}\rangle-\Omega
=\frac{1}{t}\bigl(\Omega t+\delta t+2\delta\bigr)-\Omega
=\left(1+\frac{2}{t}\right)\delta,
\label{eq:bias-conv}
\end{equation}
which is independent of $\gamma$. The mean-square error then splits into the
statistical term of Eq.~\eqref{eq:mse-gamma} and the square of this bias,
\begin{equation}
E_{\mathrm{MS}}\simeq \frac{e^{2\gamma\tau}}{M t^{2}}+b^{2}
\label{eq:mse-conv}
\end{equation}

The statistical term falls as $1/M$, but the bias does not.
As $M\to\infty$ the mean-square error saturates at
\begin{equation}
E_{\mathrm{MS}}\to b^{2}=\left(1+\frac{2}{t}\right)^{2}\delta^{2}.
\label{eq:floor-conv}
\end{equation}
This floor is set by the detuning alone and limits the single-pulse protocol.

\subsection{Composite-pulse protocol}
\label{sec:composite}
We now replace each single $\pi/2$ pulse by a composite sequence designed to cancel
the bias to first order in $\delta$. We call this the composite-pulse protocol~[Fig.~\ref{fig:example}(b)].
In the exposure Hamiltonian $\tilde{H}_{\mathrm{mag},\delta}=(\Omega+\delta)\sigma_{z}/2$, the
detuning and the target field share the same $\sigma_{z}$, so the exposure splits
symmetrically. The single-pulse dynamics of Sec.~\ref{sec:conventional} then reads, to
first order in $\delta$,
\begin{align}
&\tilde{U}_{\delta}(\pi/2,0)\,e^{-i\tilde{H}_{\mathrm{mag},\delta}t}\,
\tilde{U}_{\delta}(\pi/2,\pi/2)\nonumber\\
&=\tilde{U}_{\delta}(\pi/2,0)\,e^{-i\delta\sigma_{z}t/4}\,
e^{-i\tilde{H}_{\mathrm{mag}}t}\,e^{-i\delta\sigma_{z}t/4}\,
\tilde{U}_{\delta}(\pi/2,\pi/2)\nonumber\\
&\approx\tilde{U}_{\delta}(\pi/2,0)
\left(\sigma_{0}-i\tfrac{\delta t}{4}\sigma_{z}\right)
e^{-i\tilde{H}_{\mathrm{mag}}t}
\left(\sigma_{0}-i\tfrac{\delta t}{4}\sigma_{z}\right)
\tilde{U}_{\delta}(\pi/2,\pi/2).
\label{eq:cp-split}
\end{align}

The first-order error is cancelled if the pulses are replaced according to
\begin{align}
\tilde{U}_{\delta}(\pi/2,\pi/2)&\to
\left(\sigma_{0}+i\tfrac{\delta t}{4}\sigma_{z}\right)\tilde{U}(\pi/2,\pi/2),\nonumber\\
\tilde{U}_{\delta}(\pi/2,0)&\to
\tilde{U}(\pi/2,0)\left(\sigma_{0}+i\tfrac{\delta t}{4}\sigma_{z}\right).
\label{eq:cp-requirement}
\end{align}
Each new factor $\sigma_{0}+i(\delta t/4)\sigma_{z}$ then meets the factor
$\sigma_{0}-i(\delta t/4)\sigma_{z}$ from the exposure in Eq.~\eqref{eq:cp-split}, and
their product is $\sigma_{0}+\mathcal{O}(\delta^{2})$. The dynamics reduces to the ideal
one,
\begin{equation}
\tilde{U}(\pi/2,0)\,e^{-i\tilde{H}_{\mathrm{mag}}t}\,\tilde{U}(\pi/2,\pi/2)
+\mathcal{O}(\delta^{2}),
\label{eq:cp-ideal}
\end{equation}
to first order in $\delta$.

The requirement in Eq.~\eqref{eq:cp-requirement} is realized by a three-pulse composite sequence.
The preparation rotates about the $y$ axes,
\begin{equation}
\tilde{U}(\theta_{3},\tfrac{3\pi}{2})\,\tilde{U}(\theta_{2},\tfrac{\pi}{2})\,
\tilde{U}(\theta_{1},\tfrac{3\pi}{2}),
\label{eq:cp-prep}
\end{equation}
with pulse areas
\begin{equation}
\theta_{1}=\tfrac{3\pi}{4}+A-B,\quad
\theta_{2}=2A,\quad
\theta_{3}=\tfrac{3\pi}{4}+A+B,
\label{eq:cp-areas}
\end{equation}
where $A=\arcsin[\sqrt{2\alpha^{2}-2\alpha+1}/(2\sqrt2)]$, $B=\arcsin[\alpha/\sqrt{2\alpha^{2}-2\alpha+1}]$,
and $\alpha=t/4$.
The arcsine in $A$ is real only for $t\le2(1+\sqrt{15})$, which bounds the exposure time.
The three pulses compose to a net $3\pi/2$ rotation, which equals the $\pi/2$ rotation of the
single-pulse preparation up to a global phase.
The sequence is similar to a CORPSE pulse~\cite{PhysRevA.67.042308},
although here the areas are chosen to leave the first-order term of Eq.~\eqref{eq:cp-requirement} rather than to cancel the detuning accumulated during the pulses.
The readout uses the same areas in reversed order, about the $x$ axes,
\begin{equation}
\tilde{U}(\theta_{1},\pi)\,\tilde{U}(\theta_{2},0)\,\tilde{U}(\theta_{3},\pi).
\label{eq:cp-read}
\end{equation}
Both sequences are derived in Appendix~\ref{app:composite}.

With this composite preparation and readout, the dynamics agrees with the ideal one of
Eq.~\eqref{eq:cp-ideal} to first order in $\delta$, and the linear bias is removed.
The outcome probability becomes
\begin{equation}
P_{0}^{\mathrm{CP}}=\frac{1}{2}+\frac{1}{2}e^{-\gamma\tau_{\mathrm{CP}}}\Omega t
+\mathcal{O}(\delta^{2}),
\label{eq:P0-cp}
\end{equation}
where $\tau_{\mathrm{CP}}$ is the duration of one trial.
The $\Omega t$ term is the same as in the ideal protocol, and the detuning now enters only at second order in $\delta$.
From here on we fix the exposure time to its largest allowed value $t=2(1+\sqrt{15})$,
where $\theta_{2}=\pi$ and $\theta_{1}+\theta_{3}=\tfrac{5\pi}{2}$. The trial duration is then
\begin{equation}
\tau_{\mathrm{CP}}=t+2(\theta_{1}+\theta_{2}+\theta_{3})=t+7\pi.
\label{eq:tau-cp}
\end{equation}

\section{Performance comparison}
\label{sec:performance}

\subsection{Analytical trade-off}
\label{sec:tradeoff}

We compare the single-pulse (SP) and composite-pulse (CP) protocols at a fixed total
experimental time $T$ through the mean-square error. The number of
trials is
\begin{equation}
M_{X}=\frac{T}{\tau_{X}},\qquad X\in\{\mathrm{SP},\mathrm{CP}\},
\label{eq:M-budget}
\end{equation}
where $\tau_{X}$ is the single-trial duration of protocol $X$, so a longer trial means
fewer repetitions.
We write the corresponding exposure times as $t_{\mathrm{SP}}$ and $t_{\mathrm{CP}}$.

Consider first the single-pulse protocol.
Substituting the bias of
Eq.~\eqref{eq:bias-conv} and $M_{\mathrm{SP}}=T/\tau_{\mathrm{SP}}$ into
Eq.~\eqref{eq:mse-conv}, the mean-square error is
\begin{equation}
E_{\mathrm{MS}}^{\mathrm{SP}}
\approx\frac{e^{2\gamma\tau_{\mathrm{SP}}}\,\tau_{\mathrm{SP}}}{T\,t_{\mathrm{SP}}^{2}}
+\Bigl(1+\frac{2}{t_{\mathrm{SP}}}\Bigr)^{2}\delta^{2},
\label{eq:mse-sp-T}
\end{equation}
to leading order in $\delta$ and $\Omega t$.
The first term falls as $1/T$, but the second does not: the detuning imposes a floor
independent of $T$.

The composite sequence cancels the leading-order bias.
Setting $M_{\mathrm{CP}}=T/\tau_{\mathrm{CP}}$, the mean-square error is
\begin{equation}
E_{\mathrm{MS}}^{\mathrm{CP}}
\approx\frac{e^{2\gamma\tau_{\mathrm{CP}}}\,\tau_{\mathrm{CP}}}{T\,t_{\mathrm{CP}}^{2}}.
\label{eq:mse-cp-T}
\end{equation}
Unlike Eq.~\eqref{eq:mse-sp-T}, there is no $\delta^{2}$ floor.
However, when $\tau_{\mathrm{CP}}$ exceeds $\tau_{\mathrm{SP}}$, the larger prefactor $e^{2\gamma\tau_{\mathrm{CP}}}$ raises the statistical error, and this is the price for removing the floor.

Comparing Eqs.~\eqref{eq:mse-sp-T} and~\eqref{eq:mse-cp-T}, the composite-pulse protocol is
more precise, $E_{\mathrm{MS}}^{\mathrm{CP}}<E_{\mathrm{MS}}^{\mathrm{SP}}$, once the
detuning exceeds
\begin{equation}
\delta^{\ast}=\frac{1}{1+2/t_{\mathrm{SP}}}
\sqrt{\frac{e^{2\gamma\tau_{\mathrm{CP}}}\,\tau_{\mathrm{CP}}}{T\,t_{\mathrm{CP}}^{2}}
-\frac{e^{2\gamma\tau_{\mathrm{SP}}}\,\tau_{\mathrm{SP}}}{T\,t_{\mathrm{SP}}^{2}}}.
\label{eq:delta-star}
\end{equation}
The threshold scales as $\delta^{\ast}\propto T^{-1/2}$, so a longer experiment lowers it
and widens the range of detunings over which the composite-pulse protocol wins.

\subsection{Numerical evaluation}
\label{sec:numerics}
We evaluate both protocols by Monte Carlo simulation of the full (non-perturbative) dynamics.
The parameters are $\Omega=3\times10^{-4}$, $\lambda=1$, $\gamma=0.1$, and total experimental time $T=2\times10^{8}$.
Each protocol uses its optimal exposure time from Appendix~\ref{app:optimal},
which fixes the trial duration $\tau_{X}$ and hence the number of measurements $M_{X}=T/\tau_{X}$.
This gives $M_{\mathrm{SP}}\approx2.05\times10^{7}$ and $M_{\mathrm{CP}}\approx6.30\times10^{6}$.
For each protocol we compute the exact outcome probability $P_{0}$ from the density-matrix evolution and sample
$k\sim\mathrm{Binomial}(M_{X},P_{0})$.
The empirical frequency $k/M_{X}$ then yields an estimate $\tilde{\Omega}$ through Eq.~\eqref{eq:estimator}. Running $10^{5}$ repetitions of this procedure, we obtain the root-mean-square (RMS) error in $\tilde{\Omega}$.

\begin{figure}[h]
\includegraphics[width=\columnwidth]{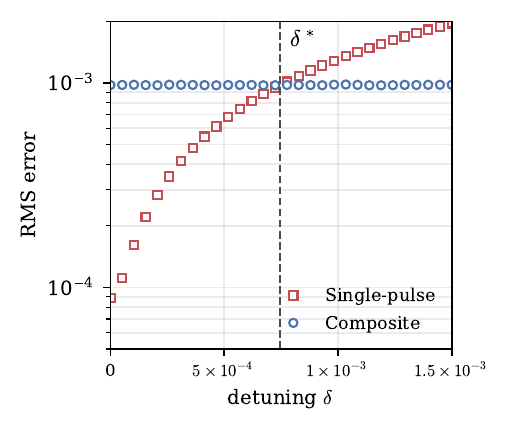}
\caption{Monte Carlo RMS error versus a detuning $\delta$ common to all measurements at
$\gamma=0.1$.
The simulation parameters are $\Omega=3\times10^{-4}$, $\lambda=1$, and total experimental time $T=2\times10^{8}$, corresponding to $M_{\mathrm{SP}}=\lfloor T/\tau_{\mathrm{SP}}\rfloor=20508008$ and
$M_{\mathrm{CP}}=\lfloor T/\tau_{\mathrm{CP}}\rfloor=6301769$, with single-pulse exposure
time $t_{\mathrm{SP}}=6.611$ and composite-pulse exposure time $t_{\mathrm{CP}}=2(1+\sqrt{15})\approx9.746$. The single-pulse error (red) rises through the bias of Eq.~\eqref{eq:bias-conv};
the composite-pulse error (blue) stays flat.
The dashed line is the analytical threshold $\delta^{\ast}\approx7.5\times10^{-4}$~[Eq.~\eqref{eq:delta-star}].}
\label{fig:rms-fixed}
\end{figure}

Figure~\ref{fig:rms-fixed} shows the RMS error against a detuning $\delta$ common to all measurements.
The single-pulse error rises with $\delta$ through the bias of Eq.~\eqref{eq:bias-conv}, whereas the composite-pulse error stays flat because the bias is removed.
The two curves cross near $\delta\approx7.5\times10^{-4}$, close to the analytical threshold $\delta^{\ast}$ of Eq.~\eqref{eq:delta-star}.
For $\delta>\delta^{\ast}$ the composite-pulse protocol has the smaller RMS error.

\begin{figure}[h]
\includegraphics[width=\columnwidth]{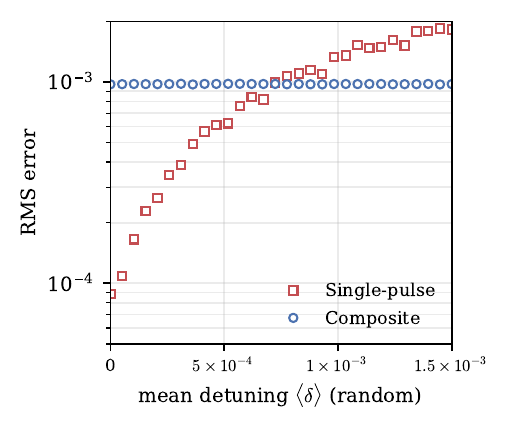}
\caption{Monte Carlo RMS error for a detuning that drifts on a long time scale at
$\gamma=0.1$. The $M_{X}$ measurements are split into $100$ blocks of
$\lfloor M_{X}/100\rfloor$ each, and each block is assigned a detuning drawn uniformly over
$[0,\delta_{\max}]$.
The horizontal axis is the mean detuning $\langle\delta\rangle=\delta_{\max}/2$.
The single-pulse error (red) rises while the composite-pulse error (blue) stays flat, and the crossover sits near the threshold of Fig.~\ref{fig:rms-fixed}.}
\label{fig:rms-drift}
\end{figure}

Figure~\ref{fig:rms-drift} treats a detuning that drifts on a long time scale rather than
staying fixed. For each protocol we split the $M_{X}$ measurements into $100$ blocks of
$\lfloor M_{X}/100\rfloor$ each and assign each block a detuning drawn uniformly over $[0,\delta_{\max}]$.
The RMS error is plotted against the mean detuning $\langle\delta\rangle=\delta_{\max}/2$.
The single-pulse error again rises while the composite-pulse error stays flat,
and the crossover sits near the threshold of Fig.~\ref{fig:rms-fixed}.
The composite-pulse advantage thus survives a slowly drifting detuning.

\section{Summary and discussion}
\label{sec:summary}
We have analyzed the detuning-induced systematic error in single-qubit magnetic-field
sensing with depolarizing noise. The detuning biases the estimator and imposes an error
floor that does not vanish as the number of measurements grows. To remove this floor, we
designed a composite-pulse preparation and readout that cancels the detuning to first
order. We then evaluated the performance of this composite-pulse protocol.

Several directions remain open. Our composite sequence cancels the detuning only to first
order, and the residual second-order terms remain to be quantified. Our scheme acts on the
control before measurement, whereas error-mitigation methods post-process the data
afterwards~\cite{PhysRevLett.129.250503,PhysRevA.109.022410,kwon2025virtual}; the two
approaches are therefore complementary. Similar composite-pulse strategies may address
other coherent control errors.

S. Kukita and K. Nawata contributed equally to this paper.
This project is supported by
    JST Moonshot R\&D Grant
    Number JPMJMS226C, 
    JST CREST Grant Number JPMJCR23I5, and Presto
    JST Grant Number JPMJPR245B.

\appendix
\section{Pulse model and rotating-wave approximation}
\label{app:rwa}
Here we derive the single-qubit propagator of Eq.~\eqref{eq:rotation}. The bare
Hamiltonian is $H_{0}=(\omega_{0}/2)\sigma_{z}$, and the drive adds
$H_{1}(s)=(\Lambda/2)\cos(\omega s+\phi)\sigma_{x}$. We move to the frame rotating at
$\omega_{0}$ through
\begin{equation}
U_{0}(s)=e^{-i\omega_{0}s\sigma_{z}/2},
\end{equation}
and factor the full propagator as
\begin{equation}
U(s)=U_{0}(s)\,\tilde{U}(s).
\end{equation}
The rotating-frame propagator $\tilde{U}$ obeys
\begin{equation}
\dot{\tilde{U}}=-i\tilde{H}_{1}\tilde{U},\qquad
\tilde{H}_{1}=U_{0}^{\dagger}H_{1}U_{0},
\end{equation}
where $H_{0}$ has been cancelled by the transformation. On resonance ($\omega=\omega_{0}$),
\begin{align}
\tilde{H}_{1}(s)&=\tfrac{\Lambda}{4}\bigl(\cos\phi\,\sigma_{x}+\sin\phi\,\sigma_{y}\bigr)\nonumber\\
&\quad+\tfrac{\Lambda}{4}\bigl[\cos(2\omega_{0}s+\phi)\sigma_{x}
-\sin(2\omega_{0}s+\phi)\sigma_{y}\bigr].
\end{align}

In the regime $\omega_{0}\gg\Lambda$ the counter-rotating terms oscillating at
$2\omega_{0}$ give only a correction of order $\Lambda/\omega_{0}$ to $\tilde{U}$. We
neglect this correction; this is the rotating-wave approximation. The effective generator
is then time independent,
\begin{equation}
\tilde{H}_{1}\approx\frac{\Lambda}{4}\,\vec{n}_{\phi}\cdot\vec{\sigma},\qquad
\vec{n}_{\phi}=(\cos\phi,\sin\phi,0).
\end{equation}
Solving the equation of motion gives Eq.~\eqref{eq:rotation}. The pulse strength there is
$\lambda=\Lambda/2$. The factor $U_{0}(s)$ is absorbed into the measurement basis and
omitted.

\section{Derivation of the tailored composite sequence}
\label{app:composite}

We derive the pulse areas $\theta_{1,2,3}$ of the composite preparation in
Sec.~\ref{sec:composite}. They follow from the requirement
\begin{align}
\tilde{U}_{\delta}(\theta_{3},\tfrac{3\pi}{2})\,&\tilde{U}_{\delta}(\theta_{2},\tfrac{\pi}{2})\,
\tilde{U}_{\delta}(\theta_{1},\tfrac{3\pi}{2})\nonumber\\
&\approx\Bigl(\sigma_{0}+i\tfrac{\delta t}{4}\sigma_{z}\Bigr)
\tilde{U}(\tfrac{3\pi}{2},\tfrac{3\pi}{2})+\mathcal{O}(\delta^{2}),
\label{eq:appB-goal}
\end{align}
which implements the first line of Eq.~\eqref{eq:cp-requirement}.

We abbreviate $U_{1}=\tilde{U}(\theta_{1},\tfrac{3\pi}{2})$,
$U_{2}=\tilde{U}(\theta_{2},\tfrac{\pi}{2})$, $U_{3}=\tilde{U}(\theta_{3},\tfrac{3\pi}{2})$.
The first-order expansion $\tilde{U}_{\delta}=\tilde{U}-i\delta\sin(\theta/2)\sigma_{z}$ of
Eq.~\eqref{eq:Udelta} turns the left-hand side into
\begin{align}
U_{3}U_{2}U_{1}-i\delta\Bigl[&\sin\tfrac{\theta_{1}}{2}\,U_{3}U_{2}\sigma_{z}
+\sin\tfrac{\theta_{2}}{2}\,U_{3}\sigma_{z}U_{1}\nonumber\\
&+\sin\tfrac{\theta_{3}}{2}\,\sigma_{z}U_{2}U_{1}\Bigr]+\mathcal{O}(\delta^{2}).
\label{eq:appB-expand}
\end{align}

Each pulse axis lies in the equatorial plane, so
$\sigma_{z}\tilde{U}(\theta,\phi)=\tilde{U}^{\dagger}(\theta,\phi)\sigma_{z}$. This moves
every $\sigma_{z}$ to the left, and the bracketed first-order term becomes
\begin{equation}
\sigma_{z}\Bigl[\sin\tfrac{\theta_{1}}{2}\,U_{3}^{\dagger}U_{2}^{\dagger}
+\sin\tfrac{\theta_{2}}{2}\,U_{3}^{\dagger}U_{1}
+\sin\tfrac{\theta_{3}}{2}\,U_{2}U_{1}\Bigr].
\end{equation}

The operators $U_{1,2,3}$ are rotations about $\pm y$, so
\begin{equation}
U_{3}^{\dagger}U_{2}^{\dagger}=\cos\tfrac{\theta_{2}-\theta_{3}}{2}\,\sigma_{0}
+i\sin\tfrac{\theta_{2}-\theta_{3}}{2}\,\sigma_{y}.
\end{equation}
The other two products, $U_{3}^{\dagger}U_{1}$ and $U_{2}U_{1}$, take the same form.
Substituting these into the bracketed term and using the product-to-sum identities,
\begin{align}
&\sin\tfrac{\theta_{1}}{2}\,U_{3}^{\dagger}U_{2}^{\dagger}
+\sin\tfrac{\theta_{2}}{2}\,U_{3}^{\dagger}U_{1}
+\sin\tfrac{\theta_{3}}{2}\,U_{2}U_{1}
=S\,\sigma_{0}+iC\,\sigma_{y},
\label{eq:appB-SC}
\end{align}
with
\begin{align}
S&=\sin\tfrac{\theta_{1}+\theta_{3}-\theta_{2}}{2}
+\sin\tfrac{\theta_{1}+\theta_{2}-\theta_{3}}{2}
+\sin\tfrac{\theta_{2}+\theta_{3}-\theta_{1}}{2},\label{eq:appB-Sdef}\\
C&=\cos\tfrac{\theta_{2}+\theta_{3}-\theta_{1}}{2}
-\cos\tfrac{\theta_{1}+\theta_{2}-\theta_{3}}{2}.\label{eq:appB-Cdef}
\end{align}

Substituting Eq.~\eqref{eq:appB-SC} into Eq.~\eqref{eq:appB-expand}, the left-hand side of
Eq.~\eqref{eq:appB-goal} becomes
\begin{equation}
U_{3}U_{2}U_{1}-i\delta\,\sigma_{z}(S\sigma_{0}+iC\sigma_{y})+\mathcal{O}(\delta^{2}).
\end{equation}
We match this with the right-hand side of Eq.~\eqref{eq:appB-goal} order by order. The
product $U_{3}U_{2}U_{1}$ is a $y$ rotation through $\theta_{1}-\theta_{2}+\theta_{3}$. The
zeroth order $U_{3}U_{2}U_{1}=\tilde{U}(\tfrac{3\pi}{2},\tfrac{3\pi}{2})$ fixes this angle
to $\tfrac{3\pi}{2}$. The first order gives
\begin{equation}
S\sigma_{0}+iC\sigma_{y}=-\tfrac{t}{4}\,\tilde{U}(\tfrac{3\pi}{2},\tfrac{3\pi}{2}).
\end{equation}
With $\tilde{U}(\tfrac{3\pi}{2},\tfrac{3\pi}{2})=-\tfrac{1}{\sqrt2}\sigma_{0}
+\tfrac{i}{\sqrt2}\sigma_{y}$ and $\alpha:= t/4$, we obtain
\begin{equation}
\theta_{1}+\theta_{3}-\theta_{2}=\tfrac{3\pi}{2},\qquad
S=\tfrac{\alpha}{\sqrt2},\qquad C=-\tfrac{\alpha}{\sqrt2}.
\label{eq:appB-cond}
\end{equation}

The zeroth-order condition $\theta_{1}+\theta_{3}-\theta_{2}=\tfrac{3\pi}{2}$ fixes the
first term of $S$ in Eq.~\eqref{eq:appB-Sdef},
\begin{equation}
\sin\tfrac{\theta_{1}+\theta_{3}-\theta_{2}}{2}=\sin\tfrac{3\pi}{4}=\tfrac{1}{\sqrt2}.
\end{equation}
The combinations
\begin{equation}
X:=\tfrac{\theta_{2}+\theta_{3}-\theta_{1}}{2},\qquad
Y:=\tfrac{\theta_{1}+\theta_{2}-\theta_{3}}{2}
\end{equation}
bring $S$ and $C$ to
\begin{equation}
S=\tfrac{1}{\sqrt2}+\sin X+\sin Y,\qquad C=\cos X-\cos Y.
\end{equation}
The conditions $S=\tfrac{\alpha}{\sqrt2}$ and $C=-\tfrac{\alpha}{\sqrt2}$ then read
\begin{align}
\sin X+\sin Y&=\tfrac{\alpha-1}{\sqrt2},\\
\cos X-\cos Y&=-\tfrac{\alpha}{\sqrt2}.
\end{align}
The sum-to-product identities turn these into
\begin{align}
2\sin\tfrac{X+Y}{2}\cos\tfrac{X-Y}{2}&=\tfrac{\alpha-1}{\sqrt2},\label{eq:appB-cos}\\
2\sin\tfrac{X+Y}{2}\sin\tfrac{X-Y}{2}&=\tfrac{\alpha}{\sqrt2}.\label{eq:appB-sin}
\end{align}

Dividing Eq.~\eqref{eq:appB-sin} by Eq.~\eqref{eq:appB-cos} gives
\begin{equation}
\tan\tfrac{X-Y}{2}=\frac{\alpha}{\alpha-1},
\end{equation}
and squaring and adding them gives
\begin{equation}
\sin\tfrac{X+Y}{2}=\pm\frac{\sqrt{2\alpha^{2}-2\alpha+1}}{2\sqrt2}.
\end{equation}
We take the positive branch, which is the one relevant to the value $\alpha>1$ used below, and define
\begin{align}
A:=&\arcsin\frac{\sqrt{2\alpha^{2}-2\alpha+1}}{2\sqrt2},\nonumber\\
B:=&\arcsin\frac{\alpha}{\sqrt{2\alpha^{2}-2\alpha+1}},
\end{align}
the first equal to $\tfrac{X+Y}{2}$ and the second to $\tfrac{X-Y}{2}$.

From $A=\tfrac{X+Y}{2}$ and $B=\tfrac{X-Y}{2}$, together with
$\theta_{1}+\theta_{3}-\theta_{2}=\tfrac{3\pi}{2}$, the pulse areas are
\begin{equation}
\theta_{1}=\tfrac{3\pi}{4}+A-B,\qquad
\theta_{2}=2A,\qquad
\theta_{3}=\tfrac{3\pi}{4}+A+B.
\label{eq:appB-theta}
\end{equation}
Here $A$ is real only for
\begin{equation}
\alpha\le\tfrac{1+\sqrt{15}}{2},
\end{equation}
equivalently $t\le2(1+\sqrt{15})$. We take the largest value,
$\alpha=\tfrac{1+\sqrt{15}}{2}$. An analogous derivation for the readout, whose pulses are
rotations about $\pm x$, gives
$\tilde{U}(\theta_{1},\pi)\tilde{U}(\theta_{2},0)\tilde{U}(\theta_{3},\pi)$, as in
Eq.~\eqref{eq:cp-read}.

\section{Optimal exposure times}
\label{app:optimal}

The exposure times $t_{\mathrm{SP}}$ and $t_{\mathrm{CP}}$ are free parameters of the two
protocols.
As the detuning $\delta$ is unknown, we cannot minimize the bias term, and we
instead choose each exposure to minimize the statistical error at fixed total time $T$. The
statistical term of Eq.~\eqref{eq:mse-gamma}, with $M_{X}=T/\tau_{X}$ from
Eq.~\eqref{eq:M-budget}, is
\begin{equation}
\frac{e^{2\gamma\tau_{X}}\,\tau_{X}}{T\,t_{X}^{2}},\qquad X\in\{\mathrm{SP},\mathrm{CP}\},
\end{equation}
which we minimize over the exposure of each protocol.

In both protocols, the pulse durations scale as $1/\lambda$ with the strength $\lambda$, which is limited by the hardware. We set this experimental limit to $\lambda=1$,
and can take $\lambda<1$ if necessary. We optimize the exposure times under this constraint.


For the single-pulse protocol, the choice of $\lambda$ is simple. The exposure $t_{\mathrm{SP}}$
does not depend on $\lambda$, while the pulses shorten as $\lambda$ grows, so $\lambda=1$ is
optimal. The trial duration is then $\tau_{\mathrm{SP}}=t_{\mathrm{SP}}+\pi$, and the
statistical term becomes
\begin{equation}
\frac{e^{2\gamma(t_{\mathrm{SP}}+\pi)}(t_{\mathrm{SP}}+\pi)}{T\,t_{\mathrm{SP}}^{2}}.
\end{equation}
Setting its derivative with respect to $t_{\mathrm{SP}}$ to zero gives
\begin{equation}
2\gamma\,t_{\mathrm{SP}}^{2}+(2\gamma\pi-1)\,t_{\mathrm{SP}}-2\pi=0,
\end{equation}
with roots
\begin{equation}
t_{\mathrm{SP}}=\frac{1-2\gamma\pi\pm\sqrt{4\gamma^{2}\pi^{2}+12\gamma\pi+1}}{4\gamma}.
\end{equation}
The root with the minus sign is negative, so we take
\begin{equation}
t_{\mathrm{SP}}^{\mathrm{opt}}
=\frac{1-2\gamma\pi+\sqrt{4\gamma^{2}\pi^{2}+12\gamma\pi+1}}{4\gamma}.
\label{eq:appC-tsp}
\end{equation}

For the composite-pulse protocol, the exposure time is set to be $t_{\mathrm{CP}}=2(1+\sqrt{15})$ when $\lambda=1$, but we can tune this by changing $\lambda$.
The exposure scales as $t_{\mathrm{CP}}=2(1+\sqrt{15})/\lambda$, and we optimize it by
varying $\lambda$. Lowering $\lambda$ lengthens the exposure but also lengthens the control,
making the optimization nontrivial.
As $\lambda=1$ is the maximum, the exposure obeys $t_{\mathrm{CP}}\ge2(1+\sqrt{15})$.
The trial duration scales with it, $\tau_{\mathrm{CP}}=\beta\,t_{\mathrm{CP}}$, where
\begin{equation}
\beta=\frac{2(1+\sqrt{15})+7\pi}{2(1+\sqrt{15})}.
\end{equation}
This gives $\tau_{\mathrm{CP}}=t_{\mathrm{CP}}+7\pi$ at $\lambda=1$, consistent with
Eq.~\eqref{eq:tau-cp}.
The statistical term is then
\begin{equation}
\frac{\beta\,e^{2\gamma\beta t_{\mathrm{CP}}}}{T\,t^{2}_{\mathrm{CP}}},
\end{equation}
and its derivative vanishes at
\begin{equation}
t_{\mathrm{CP}}^{\ast}=\frac{1}{2\beta\gamma}
=\frac{1+\sqrt{15}}{\gamma\bigl(2(1+\sqrt{15})+7\pi\bigr)},
\end{equation}
corresponding to $\lambda^{\ast}=2\gamma\bigl(2(1+\sqrt{15})+7\pi\bigr)$.

This stationary point is a minimum, but the exposure must satisfy
$t_{\mathrm{CP}}\ge2(1+\sqrt{15})$, corresponding to $\lambda\le 1$.
When $t_{\mathrm{CP}}^{\ast}$ falls below this bound, the
minimum over the allowed range is attained at the bound. The optimal exposure is therefore
\begin{equation}
t_{\mathrm{CP}}^{\mathrm{opt}}
=\max\!\left(\frac{1+\sqrt{15}}{\gamma\bigl(2(1+\sqrt{15})+7\pi\bigr)},\;2(1+\sqrt{15})\right).
\label{eq:appC-tcp}
\end{equation}
At the value $\gamma=0.1$ used in the main text, the first argument is smaller than
$2(1+\sqrt{15})$, so the optimum is $t_{\mathrm{CP}}^{\mathrm{opt}}=2(1+\sqrt{15})$, attained
at $\lambda=1$.
Both protocols therefore use $\lambda=1$ in the numerical study of Sec.~\ref{sec:numerics}.

\bibliography{sensing}

\end{document}